# New Theoretical Insights Unraveling Color Pattern in the Flowers of *Passiflora incarnata*

Ishaan Misra and V. Ramanathan*
Department of Chemistry, Indian Institute of Technology (BHU), Varanasi
(*Author to whom co-respondence must be addressed at vraman.chy@iitbhu.ac.in)

Abstract: The change in the color pattern of the petals of *Passiflora incarnata* is studied using the chaos theory in the form of logistic maps and plotted using the corresponding bifurcation diagram. Based on a colorful inspection of the beginning of violet-colored dots along the filament of the flower's bud stage and the emergence of alternating bands of violet and white color in the matured bloom, it is possible to deduce that a two-degree model for polynomial mapping can be used to study color oscillations in the flower.

## 1.1 Introduction

Appealing symmetry and patterns abound in the biological world[1]. These can resemble variously shaped organisms or have colourful patterns on their surfaces or skin. The appearance of such patterns in nature is of great interest, both for understanding a range of phenomena in natural and artificial systems, as well as for the specific situations in which such patterns are discovered. It is uncertain whether a pattern fully develops and at what point in an organism's growth it starts to show up. Turing proposed a mathematical model as part of the reaction-diffusion (RD) theory to demonstrate that morphogenesis could lead to the formation of spatial patterns triggered by random fluctuations [2]. A more general form of the reaction-diffusion equation, derived from its original form proposed by Turing [3] is

$$\frac{\partial u}{\partial t} = D\nabla^2 u + f(u,p) \quad (1)$$

where u is a vector of chemical concentrations, D is a matrix of diffusion co-efficient and f is the chemical coupling matrix with kinetic parameters p. Considering only two variables A and I to represent the activator and inhibitor in equation(1), we obtain

$$\frac{\partial I}{\partial t} = D_I \nabla^2 I + f(A,I) \quad (2)$$

$$\frac{\partial A}{\partial t} = D_A \nabla^2 A + f(A,I) \quad (3)$$

Various studies have been done to explain temporal oscillations in the Passion flower, *Passiflora incarnata*(PI)*,* starting from the modified model[4]. The model comprised a florigen, that is responsible for flowering and an antigen that works against flowering[5,6]. Although the florigen and the anti-florigen responsible for the temporal oscillations in PI flowering remain to be identified in biochemical terms, the former plays an autocatalytic role and the latter an inhibitory role. Sathyamurthy and co-workers used the design of an activator-inhibitor model to understand the underlying principles in the formation of alternating violet and white bands in the PI flower where they adopted the Gierer-Meinhardt kinetics and formulated the following equations[7].

$$\frac{\partial A}{\partial t} = \frac{\rho A^2}{I} - \mu A + D_A \frac{\partial^2 A}{\partial x^2} + \rho_o \quad (4a)$$

$$\frac{\partial I}{\partial t} = \rho' A^2 - \nu I + D_I \frac{\partial^2 I}{\partial x^2} \quad (4b)$$

where A is the activator concentration, I is the inhibitor concentration, $\rho$, $\mu$ and $\nu$ are first order rate coefficients, $\rho_o$ is a second order rate coefficient, $D_A$ and $D_I$ are diffusion coefficients of the activator and the inhibitor, respectively. Equations (4a) and (4b) on solving give the spatio-temporal evolution of A and I[7].

In this work, we have applied the logistic map model to linearly map the oscillatory violet and white colour pattern in the PI flower in an effort to reduce the parametric dependence of pattern formation to distance from the base of the filament only. A logistic map is a polynomial mapping of degree 2, often used to understand complex, chaotic behaviour arising from simple non-linear dynamical equations. Particularly, we look at the bifurcation diagram of a logistic map that shows the values approached asymptotically as a function of the bifurcation parameter in the system.

## 1.2 Bifurcation Diagram-Logistic Map Model

Bifurcation diagrams and logistic maps have been used to study chaos patterns arising from unstable conditions in activator-inhibitor type models like Gierer-Meinhardt model or chaos arising as a result of saturation of activator or inhibitor[8,9,10]. Logistic maps are mathematical functions that describe the population dynamics of a species in a given environment. They are commonly used to model the growth of a population over time, taking into account factors such as reproduction rates and resource availability. The logistic map is a specific form of a recurrence relation, often used in the context of discrete-time dynamical systems. Mathematically the logistic map model can be written as

$$x_n = r x_n (1 - x_n) \quad (5)$$

where:
- $x_n$ is the population at time n,
- $x_n+1$ is the population at the next time step n+1,
- r is a parameter that represents the reproductive rate or the strength of feedback in the system.

The bifurcation diagram is a powerful tool to explore the behaviour of dynamic systems and is widely used in the study of chaos theory and nonlinear dynamics like logistic maps. It provides insights into the complexity and unpredictability that can arise in seemingly simple mathematical models.

In a bifurcation diagram for a logistic map, r is chosen between 0 and 4. When r is varied between 0 and 4, the logistic map exhibits a variety of bifurcation patterns, including period-doubling bifurcations and chaotic behaviour. As r increases within this range, the system undergoes transitions from stable fixed points to periodic oscillations, and eventually to chaotic behaviour. The Python code (shown in Annexure 1 in the Supporting Information) was developed to demonstrate a general bifurcation diagram for a logistic map that can be run on any system. It uses a system of two loops to iterate through values of r from 1 to 4 with measurements taken at every 1/2000$^{th}$ interval and find the population at each value of r.

Now, if we look at the bifurcation diagram generated by the code (Annexure 1) we observe a repetitive colour pattern in the chaos regions which can be mapped to the development of the PI bud.

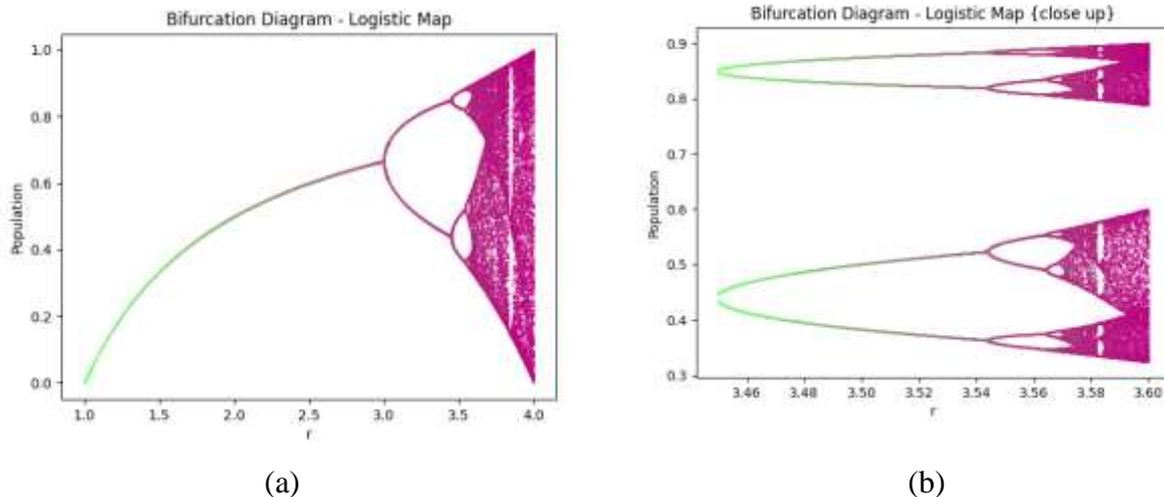

(a) (b)

**Figure 1.(a)** General representation of a bifurcation diagram of a logistic map.**(b)** Close-up of a bifurcation diagram of a logistic map : If we zoom into the bifurcation diagram by changing the range of the parameter r from [1,4] to [3.45,3.6] we get the following pattern.

From Figures (1) and (1b) we can observe that the more we zoom into the graph the more we observe the same bifurcation pattern occurring multiple number of times leading up to chaos. This feature renders the simulation of repetitive patterns like the activator-inhibitor model (filament of PI flower) much easier on the logistic map model. Table 1 shows data taken from Bhati et al's paper which serves as the basis of our mapping. [7]

**Table 1.** Measured distance of the first 4 violet bands from the centre of the flower to the base of the filament.

| Color of circle | Distance from center (cm) |
|---|---|
| V1 | 0.61 |
| V2 | 0.73 |
| V3 | 0.88 |
| V4 | 1.12 |

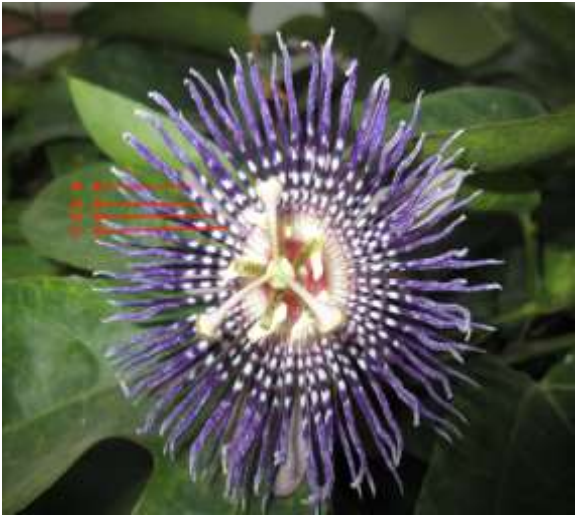

**Figure 2.** Color of circle as shown in Table 1 depicted in the top-view of the PI flower[7]

We tweak our original model to
$$x_n = (r - 0.075)x_n(1 - x_n) \tag{6}$$

This offset of 0.075 in 'r' is done manually to optimize the logistic map model to efficiently and accurately map the color bands in the filament. This model considers the first 4 violet bands (Table 1) by visualizing the bifurcation diagram as a model to represent the colour pattern in the filaments of the flower given that the colour pattern in the flower represents a non-linear dynamical system. On adjusting the widths of the chaotic region to match the violet patch of the small filament (filament in its early stages of development) in the bifurcation diagram, we can extend the logistic map with the same constraints to predict the filament growth and colour pattern. We use the code as shown in Annexure 2 (see Supporting Information) to create a bifurcation map model to map a filament of the PI flower. And measure distances along the length of the PI filament(x-axis) to get the following graph. The code shown in Annexure 2 is similar to that of Annexure 1 with modifications made to the original logistic map equation. It uses the function *logistic_map()* to generate the bifurcation diagram as shown in Figure 3b.

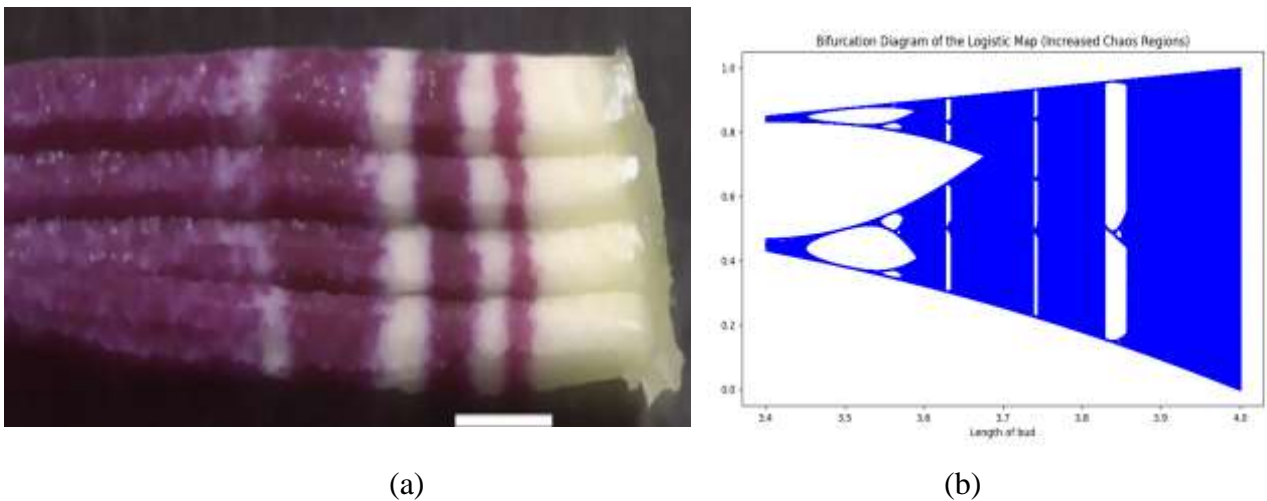

(a)  (b)

**Figure 3.** Bifurcation diagram mapped to a PI filament. **(a)** close up view of the alternating white (at the base of the filament) and violet color bands along the filaments in a matured flower[7] **(b)** Bifurcation diagram of a logistic map to represent the colour pattern in PI flower.

The chaos region is the width of the color region in the bifurcation diagram, and it is calculated based on the distance of the region from the base to the average centre of the chaos region to mimic the distance calculated from the base of the filament to the violet color band.

We calculate the location of the chaotic region as shown in Figure 4 which resembles the first 4 violet and white colour patterns while ignoring any inconsistencies or connections between the chaotic regions. In V1(violet band region 1), although chaos exists before the start of the red line we consider the minimum regions where chaos exists indicating the presence of one color (violet) without any residual white presence. We consider an arbitrary value for the base of the filament at 3.4 units and calculate the distances using the Lyapunov exponents.

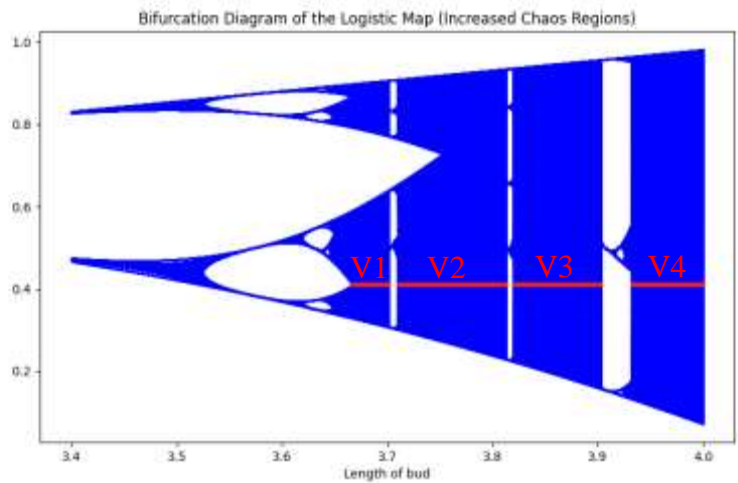

**Figure 4.** Red line marks the chaotic regions under consideration

### 1.3 Lyapunov Exponent

Lyapunov exponent ($\lambda$) measures the rate of exponential divergence or convergence of nearby trajectories in a dynamical system. In a chaotic system, small differences in the initial conditions can lead to vastly different outcomes over time. The Lyapunov exponent provides a numerical measure of this sensitivity. A positive Lyapunov exponent indicates chaotic behaviour, as trajectories diverge over time, whil2e a negative Lyapunov exponent suggests convergence. Mathematically, for a system described by a set of differential equations, the Lyapunov exponent is often defined as the average rate of exponential growth or decay of the separation between nearby trajectories. The Lyapunov exponent is a crucial tool in studying and characterizing chaotic behaviour in dynamical systems[11,12,13].

Lyapunov exponent was used to find the width and location of the color bands. Location in this context means the distance of the chaos region from the base to the average center of the chaos region to mimic the distance calculated from the base of the filament to the violet color band. Among many uses of Lyapunov exponent, its properties of identifying critical points where bifurcations occur, was used and a quantitative measurement of chaos was performed.

We use the code as provided in Annexure 3 (see Supporting Information) to calculate the Lyapunov exponent of our model and determine the points that we consider as the start(S) and end(E) of the color bands in Table 2. This Python code calculates the Lyapunov exponent for a logistic map over a specified range of bifurcation parameters (r) between 3.4 and 4.0. It then identifies major chaotic regions within this range and computes their respective widths, presenting both a plot highlighting chaotic regions and a summary of their widths. The code in Annexure 3 gives results in a plot shown in figure 5.

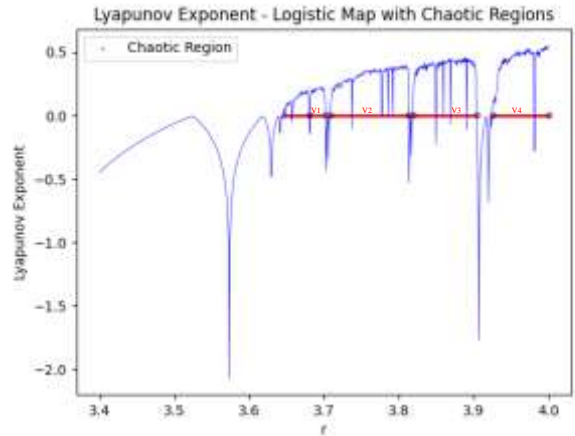

**Figure 5.** Lyapunov exponent for given r

**Table 2.** Measured distance of the first 4 blue chaos bands from the base.

| Color of circle | Chaos Region **Start(S)** | Chaos Region **End(E)** | Location = {S+E}/2 | Distance of base to chaos region (visual units) |
|---|---|---|---|---|
| V1 | 3.68169 | 3.70157 | 0.29163 | 0.47808 |
| V2 | 3.70814 | 3.81265 | 0.360395 | 0.49369 |
| V3 | 3.81805 | 3.90333 | 0.46069 | 0.52351 |
| V4 | 3.931 | 4 | 0.56249 | 0.50222 |

With a scaling factor of 2.004, the values we obtained from our model in Table 2 (column 5), match exceedingly well with real values i.e. Table 1 i.e. further calculation from our model should be scaled by a factor of 2 to maintain coherence in data.

### 1.4 Observations

Although the pattern evolving from our calculations is not exactly the same as observed in nature because of the difference in measurement units, it appears that the logistic map model theory could account for the pattern in PI if appropriately scaled.
Now that we have the distance of the chaos regions from the base in Table 2 (column 5) and the distances of the violet bands from the centre in Table 1, we need to observe if scaling the values from our observation by a common factor gives us the required values in Table 1. Dividing the values of Table 1 by Table 2(column 5), we get the scaling parameters for the four bands.

**Table 3.** Scaling parameter.

| Color of circle | Scaling parameter |
|---|---|
| V1 | 2.09169 |
| V2 | 2.02555 |
| V3 | 1.91017 |
| V4 | 1.99112 |
| **Average = 2.00463** | |

This implies that the values observed from our model are scaled down by a factor of approximately 2 for each violet band from the PI model. We calculate the standard deviation of the scaling parameter for all violet bands in Table 3 which comes out to be 0.0654. This value is in the range of 1e-2 signifying that the values are tightly wound around the average(having less spread). Therefore we can consider 2.004 as our scaling parameter for our logistic map model to represent the PI flower filament i.e. further calculation from our model should be scaled by a factor of 2 to maintain coherence in data. This table strengthens logistic maps as a model for activator-inhibitor systems by not only mapping the model as a whole but also mapping all individual chaos regions to the violet bands of the filament.

### 1.5 Conclusion

This article explores the possibility of using logistic maps as an alternative to the oscillatory equations predominantly used for activator-inhibitor-type systems. It is hereby conclusively demonstrated that the open-source Python-based logistic map model works well for a small filament in a PI flower. Extension of the current study would be rewarding because the bifurcation diagram of a logistic map is fascinating and rich in its displayed behaviour. These pieces of code are general. With a little tweak in the structures of the equation and the parameters involved graph for the desired non-linear dynamical system can be produced.
While logistic maps are highly idealized models that may not capture all the complexities of the real world. Due to the qualitative nature of this study, our model describes a phenomenological description of dynamical behaviour without diving into its predictive power or the biological mechanisms at play.
An interesting project that can be derived from this paper is to generalize this piece of code for other oscillatory systems to observe various reactions. The qualitative characteristics of the presented logistic map model can be used to model other activator-inhibitor systems by tuning the single parameter it uses. Repetitive or damped oscillatory patterns in both space and time domain can be modelled using this method due to the presence of smaller alternative chaos patterns which can be observed on zooming in the bifurcation graphs in the chaos regions. Since our model uses only a single parameter it is easy to investigate how change in parameter affects the dynamics of the system.

## 1.6 Acknowledgement
The authors thank the academic infra structure of IIT(BHU) Varanasi.

## 1.7 Declarations:
The authors declare absolutely no conflict of interests.

## 1.8 Author Contributions
VR conceptualized the idea, IM wrote and executed the python code. IM and VR together wrote the draft.

## 1.9 References

1. Ball P 2016 Patterns in nature: why the natural world looks like the way it does (Chicago: University of Chicago Press)

2. Turing AM 1952 The chemical basis of morphogenesis. Phil. Trans. R. Soc. B 237 37–72.

3. Murray JD 2003 Mathematical biology 3rd ed., volume II (Berlin: Springer)

4. Lotka AJ 1910 Contributions to the theory of periodic reactions. J. Phys. Chem. 14 271–274

5. Chailakhyan MK 1936 New facts in support of the hormonal theory of plant development. Dokl. Akad. Nauk SSSR 4 77–81

6. Lang A and Melchers G 1943 Die photoperiodische reaktion von Hyoscyamus niger. Planta 33 653–702

7. Bhati AP, Goyal S, Yadav R, et al. 2021 Pattern formation in Passiflora incarnata: An activator-inhibitor model. J. Biosci. 46 84

8. Gu L, Gong P., Wang H. 2020 "Hopf bifurcation and turing instability analysis for the gierer-meinhardt model of the depletion type," Discrete Dynamics in Nature and Society, vol. 2020, Article ID 5293748, 10 pages.

9. Li Y., Zhong S., Hou X., Wang J. et al. 2019 Analysis of bifurcation, chaos and pattern formation in a discrete time and space Gierer Meinhardt system Chaos, Solitons & Fractals, Elsevier, vol. 118(C), pages 1-17.

10. Shanshan Chen, Junping Shi & Junjie Wei (2014) Bifurcation analysis of the Gierer–Meinhardt system with a saturation in the activator production, Applicable Analysis: An International Journal, 93:6.

11. Rana, Sarker 2019 Dynamics and chaos control in a discrete-time ratio-dependent Holling-Tanner model. J. of the Egyptian Mathematical Society. 27. 10.1186/s42787-019-0055-4.

12. Zhao, J. 2020 Complexity and chaos control in a discrete-time Lotka–Volterra predator–prey system. J. Difference Equations and Applications, 1–18.

13. Santra P.K., Mahapatra G. S., Phaijoo G. R. 2020 Bifurcation and Chaos of a Discrete Predator-Prey Model with Crowley–Martin Functional Response Incorporating Proportional Prey Refuge Mathematical Problems in Engineering, vol. Article ID 5309814, 18 pages

# Supporting Information

New Theoretical Insights Unraveling Color Pattern in the Flowers of *Passiflora incarnata*

Ishaan Misra and V. Ramanathan*
Department of Chemistry, Indian Institute of Technology (IIT-BHU), Varanasi
(*Author to whom co-respondence must be addressed at vraman.chy@iitbhu.ac.in)

## Annexure 1

```python
import numpy as np
import matplotlib.pyplot as plt
rs=np.linspace(1,4,2000)
N=500
x=.5+np.zeros(N)
endcap=np.arange(round(N*.9),N)
for ri in range(len(rs)):
    for n in range(N-1):
        x[n+1]=rs[ri]*x[n]*(1-x[n])
    u=np.unique(x[endcap])
    r=rs[ri]*np.ones(len(u))
    plt.xlabel('r')
    plt.ylabel('Population')
    plt.title('Bifurcation Diagram - Logistic Map')
    plt.plot(r,u,'.',markersize=1,color=[(np.sin(ri/len(rs)/2)+1)/2,1-ri/len(rs),.5])

plt.show()
```

## Annexure 2

```python
import numpy as np
import matplotlib.pyplot as plt

# Function to generate the logistic map
def logistic_map(r, x0, num_iterations):
    population = []
    x = x0
    for _ in range(num_iterations):
        x = r * x * (1 - x)
        population.append(x)
    return population

# Parameters
num_iterations = 2000  # Increase the number of iterations
num_points = 10000    # Increase the number of data points
r_values = np.linspace(3.4, 4.0, num_points)  # Expand the range of r values
x0 = 0.5  # Initial population
shift_value = 0.075  # Adjust the shift value as needed
# Generate the bifurcation diagram
bifurcation_data = []
```

```python
for r in r_values:
    population = logistic_map(r - shift_value, x0, num_iterations)
    bifurcation_data.extend([(r, x) for x in population])

# Plot the bifurcation diagram
plt.figure(figsize=(10, 6))
plt.scatter(*zip(*bifurcation_data), s=0.1, marker='.', c='blue')
plt.xlabel('Length of bud')
plt.title('Bifurcation Diagram of the Logistic Map (Increased Chaos Regions)')
plt.show()
```

**Annexure 3**

```python
import numpy as np
import matplotlib.pyplot as plt

def lyapunov_exponent(r, x, n):
    sum_lyapunov = 0.0
    for _ in range(n):
        x = r * x * (1 - x)
        sum_lyapunov += np.log(np.abs(r - 2 * r * x))
    return sum_lyapunov / n

# Define the range of r values
r_values = np.linspace(3.4, 4.0, 1000)

# Calculate the Lyapunov exponent for each r value
lyapunov_values = [lyapunov_exponent(r-0.075, 0.5, 1000) for r in r_values]

# Define a threshold for chaos detection
chaos_threshold = 0.00  # You can adjust this threshold as needed

# Find indices where Lyapunov exponent crosses the threshold
chaotic_indices = np.where(np.array(lyapunov_values) > chaos_threshold)[0]

# Find the boundaries of each chaotic region
boundaries = np.where(np.diff(chaotic_indices) > 1)[0] + 1

# Initialize lists to store chaotic regions and their widths
chaotic_regions = []
chaotic_widths = []

# Process each chaotic region separately
for i in range(len(boundaries) + 1):
    if i == 0:
        start_index = 0
    else:
        start_index = boundaries[i - 1]

    if i == len(boundaries):
```

```python
        end_index = chaotic_indices[-1] + 1
    else:
        end_index = boundaries[i]

    chaotic_region = r_values[chaotic_indices[start_index:end_index]]
    chaotic_regions.append(chaotic_region)

    # Calculate the width of each chaotic region
    chaotic_width = chaotic_region.max() - chaotic_region.min()
    chaotic_widths.append(chaotic_width)

# Plot the Lyapunov exponent with markers indicating the chaotic regions
plt.plot(r_values, lyapunov_values, 'b-', linewidth=0.5)
plt.scatter(np.concatenate(chaotic_regions), [chaos_threshold] * len(chaotic_indices), color='red', marker='.',
s=2, label='Chaotic Region')
plt.xlabel('r')
plt.ylabel('Lyapunov Exponent')
plt.title('Lyapunov Exponent - Logistic Map with Chaotic Regions')
plt.legend()
plt.show()
```